\documentclass[twoside,12pt]{article}
\usepackage{amssymb}
\newlength{\dinwidth}
\newlength{\dinmargin}
\newtheorem{theorem}{Theorem}[section]
\newtheorem{prop}{Proposition}[section]
\newtheorem{lemma}{Lemma}[section]
\newtheorem{cor}{Corollary}[section]
\newtheorem{definition}{Definition}[section]
\newenvironment{proof}{\medskip \noindent 
            {\bf Proof.}}{ \hfill $\square$ \medskip}
\def\idty{{\leavevmode\hbox{\rm 1\kern -.3em I}}}
\def\nind{\noindent}
\def\inet{{\{\As_i \}_{i \in I}}}
\def\irnet{{\{\Rs_i \}_{i \in I}}}

\def\wrnet{{\{\Rs(W)\}_{W \in\Ws}}}

\def\rnet{{\{\Rs(\Os)\}_{\Os\in\Ss}}}
\def\msc{{Modular Stability Condition}}
\def\diag{{\textnormal{diag}}}

\def\As{{\cal A}}

\def\Gs{{\cal G}}
\def\Hs{{\cal H}}

\def\Js{{\cal J}}

\def\Os{{\cal O}}
\def\Ps{{\cal P}}

\def\Rs{{\cal R}}
\def\Ss{{\cal S}}
\def\Ts{{\cal T}}

\def\Ws{{\cal W}}

\def\Zs{{\cal Z}}

\def\Pid{{\Ps_+ ^{\uparrow}}}

\def\idty{{\leavevmode\hbox{\rm 1\kern -.3em I}}}
\def\RR{\textnormal{{\rm I \hskip -5.75pt R}}}

\def\ad{{\mathrm{ad}}}
\def\diag{{\textnormal{diag}}}

\def\Pid{{\Ps_+ ^{\uparrow}}}

\begin{document}
\title{On Deriving Space--Time From Quantum Observables and States} 

\author{{\ Stephen J.\ Summers}\\
Department of Mathematics, University of Florida,\\
Gainesville FL 32611, USA\\
\vphantom{X}\\
{and Richard K. White } \\
Department of Mathematics, Edinboro University of Pennsylvania,\\
Edinboro PA 16444, USA\\}

\date{\normalsize{
Dedicated to Rudolf Haag on the occasion of his eightieth birthday
}} 

\maketitle 

\abstract{We prove that, under suitable assumptions, operationally 
motivated quantum data completely determine a space--time in which
the quantum systems can be interpreted as evolving. At the same time,
the dynamics of the quantum system is also determined. To minimize 
technical complications, this is done in the example of three-dimensional
Minkowski space.}

\section{Introduction}

     The problem of determining from which physical observations one
can derive the properties of the space--time in which the observer is
located is as old as the theory of general relativity itself, and has
received many sorts of answers. As this is not the place to give
either a review of these or a discussion of their relative advantages
and disadvantages, we shall content ourselves with referring the
interested reader to the papers \cite{EPS,Au,AL,Ban,BHF} and the
further references to be found there. Some of these approaches use
only classical objects and data as input, others use a mixture of
classical and quantum data, while yet others begin with purely quantum
data.  In this paper we propose a novel approach to the problem which
is based upon recent advances in algebraic quantum field theory.

     Over a quarter century ago, groundbreaking work \cite{BW}
revealed a connection between the Poincar\'e group and the modular
objects which Tomita-Takesaki theory \cite{Tak,BR,KR} associates with
the vacuum state on Minkowski space and certain local algebras of
observables. This has led to many unexpected applications of modular
theory to quantum field theory --- see \cite{Bor} for a recent
review. One line of research \cite{BS,Sum,BFS1,BDFS,BS2} growing out of
Bisognano and Wichmann's work has drawn attention to the group $\Js$
generated by the modular involutions $J_W$ associated to the algebras
of observables $\As(W)$ localized in the wedge--shaped regions $W$ in
Minkowski space. That program (which has been generalized to curved
space--times) will also not be reviewed here. But it is essential to
understand that the modular involutions $J_W$ are uniquely determined
by the state, which models the preparation of the quantum system, and
the algebras $\As(W)$, whose self-adjoint elements model the
observables of the system. In other words, the group $\Js$ is
determined, at least in principle, by operational quantum data. To
eliminate the tacit reference to Minkowski space in the local
algebras, and to strengthen the purely operational nature of the
initial data, we shall consider a state $\omega$ on a collection $\inet$ 
of algebras $\As_i$ indexed by ``laboratories''
$i \in I$.\footnote{More detailed definitions will be given in Section
3.} The algebra $\As_i$ will be thought of as being generated by all
the observables measurable in the laboratory $i \in I$. One then has a
collection of modular involutions $J_i$ acting on some Hilbert space,
and they generate a group $\Js$, which has been called the modular
symmetry group.

     We therefore have an abstract group $\Js$ generated by
involutions.  This is precisely the starting point of the program of
absolute geometry, see {\it e.g.} \cite{Ah,Ba,BBPW}. From such a group
and a suitable set of axioms to be satisfied by the generators of that
group, absolute geometers derive various metric spaces such as
Minkowski spaces and Euclidean spaces upon which the abstract group
acts as the isometry group of the metric space. Different sets of
axioms on the group yield different metric spaces. This affords us with
the possibility of deriving a space--time from the group $\Js$, 
{\it i.e.}, the operational data $(\omega,\inet)$ 
would determine the space--time in which the quantum systems could 
naturally be considered to be evolving. Different sets of algebraic
relations in $\Js$ would lead to different space--times.

     It is the purpose of this paper to indicate how this could be
possible. To minimize technical complications which would detract from the
point of principle we wish to make, we shall illustrate this program
using the example of three-dimensional Minkowski space. We emphasize 
that we are establishing a conceptual point of principle --- we are
not proposing a concrete operational procedure to determine space--time.

     In Section 2 we shall present the absolute geometry relevant for
our immediate purposes. Although Minkowski space is one of the cases
the absolute geometers have already treated, their tacit geometric
starting point is different from ours, and so we are obliged to
provide another chain of arguments to derive three-dimensional
Minkowski space from the group $\Js$. The quantum data which then
``determine'' three-dimensional Minkowski space will be discussed in
Section 3. It will be shown that under purely algebraic conditions on
the group $\Js$, there exists an identification of laboratories $i \in
I$ with subregions $W_i$ of $\RR^3$ such that (1) $\Js$ contains a
representation of the Poincar\'e group on $\RR^3$ and (2) the
collection $\{\As(W_i)\}_{i\in I}$ is a Poincar\'e covariant and local
net of von Neumann algebras on $\RR^3$ satisfying Haag duality. Some
further results are proven in Section 3, as well. We make some final
comments in Section 4.

\section{Absolute Geometry and Three-Dimensional Minkowski Space}

     We first provide an overview of our reasoning in this section.
Three-dimensional Minkowski space is an affine space whose plane
at infinity is a hyperbolic projective--metric plane \cite{Cox1}. In
\cite{BBPW}, Bachmann, Pejas, Wolff, and Baur (BPWB) consider an abstract
group $\mathfrak{G}$ which is generated by an invariant system $\mathcal{G}$
of generators in which each of the generators is involutory and which
satisfies a certain set of axioms. From this they construct a hyperbolic
projective--metric plane in which the given group $\mathfrak{G}$ is
isomorphic to a subgroup of the group of congruent transformations
(motions) of the projective--metric plane. By interpreting the elements
of $\mathcal{G}$ as line reflections in a hyperbolic plane,
BPWB show that the hyperbolic projective--metric plane can be generated
by these line reflections in such a way that these line reflections
form a subgroup of the group of motions of the projective--metric plane.

     Coxeter shows in \cite{Cox3} that every motion of the hyperbolic
plane is generated by a suitable product of orthogonal line
reflections, where an orthogonal line reflection is defined as a
harmonic homology with center exterior point and axis the given
ordinary line and where the center and axis are a pole-polar pair. In
the following we show that Coxeter's and BPWB's notions of motions
coincide in the hyperbolic projective--metric plane and that the
reflections can be viewed as reflections about exterior points.

     Then we embed our projective--metric plane into a
three-dimensional projective space. By singling out our original plane
as the plane at infinity, we obtain an affine space whose plane at
infinity is a hyperbolic projective--metric plane, which is the
well-known characterization of three-dimensional Minkowski space.
We show that the motions of our original plane induce motions
in the affine space and, by a suitable identification, we show that
any motion in this Minkowski space can be generated by reflections
about spacelike lines. Thus, to construct a three-dimensional
Minkowski space, one can start with a generating set $\mathcal{G}$ of
reflections about spacelike lines, equivalently, reflections about
exterior points in the hyperbolic projective--metric plane at infinity. 
This equivalence is important for our argument. We therefore
obtain a three-dimensional affine space with the Minkowski metric, which is
constructed out of a group generated by a set of isometries. 

     The approach in this paper differs from the method used both in
\cite{Wo} for two-dimensional Minkowski space and in \cite{KO} for
four-dimensional Minkowski space. In these papers one begins by
constructing the affine space. In the two-dimensional case \cite{Wo},
the elements of the generating set $\mathcal{G}$ are identified with
line reflections in an affine plane, while in the four-dimensional
case \cite{KO}, the elements of the generating set $\mathcal{G}$ are
identified with reflections about hyperplanes in an affine
space. Thus, in each of these papers, the generating set $\mathcal{G}$
is identified with a set of symmetries. A map of affine subspaces is
then obtained using the definition of orthogonality given by commuting
generators. This map induces a hyperbolic polarity in the hyperplane
at infinity, thereby yielding the Minkowski metric.

     In our approach, we take the dual view, beginning with points
instead of lines and constructing the affine space out of the plane at
infinity. The definition of orthogonality induced by the commutation
relations of the generators in the hyperplane at infinity is used to
obtain the polarity and then the hyperplane at infinity is embedded in
an affine space to get Minkowski space. This argument is necessitated
here not only because of the different dimensionality of the space,
but also because, for reasons made clear in Section 3, our generating
involutions must ultimately have the geometric interpretation of
reflections about spacelike lines.  Despite the new elements in our
approach, much of our argument in this section consists of appropriate
re-interpretations and modifications of work already in the
literature.

\subsection{Construction of $\Pi$}

     As the ideas and results of the absolute geometers are not widely
known, particularly among theoretical and mathematical physicists,
we shall give here the definitions, axioms and main results we shall
need and also provide a sketch of some of the pertinent arguments. 
For detailed proofs, the reader is referred to \cite{BBPW} or to \cite{Ba}.

     One begins with a group $\mathfrak{G}$ generated
by an invariant system $\mathcal{G}$ of involution elements. The elements 
of $\mathcal{G}$ will be denoted by lowercase Latin
letters. Those involutory elements of $\mathfrak{G}$ which can be
represented as a product $ab$, where $a,b\in \mathcal{G}$, will be denoted by
uppercase Latin letters. If $\xi ,\eta \in \mathfrak{G}$ and $\xi \eta $ is
an involution, we shall write $\xi |\eta$. The notation 
$\xi ,\eta |\varphi ,\psi $ means $\xi |\varphi $ and $\xi |\psi $ and 
$\eta |\varphi$ and $\eta |\psi$.

     The axioms we shall use to derive three-dimensional Minkowski space
from $(\mathfrak{G},\Gs)$ are:

\medskip

\noindent \textbf{Axioms}

A.1: For every $P,Q$\ there exists a\ $g$\ with $P,Q$\ $|g.$

A.2: If $P,Q$\ $|g,h$\ then $P=Q$\ or $g=h.$

A.3: If $a,b,c$\ $|P$\ then $abc=d\in \mathcal{G}.$

A.4: If $a,b,c$\ $|g$\ then $abc=d\in \mathcal{G}.$

A.5: There exist $g,h,j$\ such that $g$\ $|h$\ but $j\nmid g,h,gh.$

A.6: For each $P$ and $g$ with $P \nmid g$ there exist exactly two
distinct elements $h_1,h_2 \in \mathcal{G}$ such that 
$h_1,h_2 \mid P$ and $g,h_i \nmid R,c$ for any $R,c$ and $i=1,2$.   

\medskip

\nind We shall call a pair $(\mathfrak{G},\mathcal{G})$
consisting of a group $\mathfrak{G}$ and an invariant system
$\mathcal{G}$ of generators of the group $\mathfrak{G}$ satisfying the
axioms above a group of motions.

     In \cite{BBPW} the elements of $\mathcal{G}$ are interpreted as
secant or ordinary lines in a hyperbolic plane. In our approach, we
view the elements of $\mathcal{G}$ initially as exterior points in a
hyperbolic plane. After embedding our hyperbolic projective--metric
plane into an affine space, we will be able to identify the elements
of $\mathcal{G}$ with spacelike lines and their corresponding
reflections in a three-dimensional Minkowski space. After realizing
that statements about the geometry of the plane at infinity correspond
to statements about the geometry of the whole space where all lines
and all planes are considered through a point, we see that the axioms
are also statements about spacelike lines --- the elements of
$\mathcal{G}$ --- and timelike lines --- the elements $P$ of
$\mathfrak{G}$ --- through any point in Minkowski 3-space.

     These algebraic axioms have a geometric interpretation in the 
group plane $(\mathfrak{G},\mathcal{G})$, which we now indicate.
The elements of $\mathcal{G}$\ are called lines of the group plane, and
those involutory group elements which can be represented as the product of
two elements of $\mathcal{G}$\ are called points of the group plane. Two
lines $g$\ and $h$\ of the group plane are said to be perpendicular if 
$g \mid h$. Thus, the points are those elements of the group which can 
be written as the product of two perpendicular lines. A point $P$\ is 
incident with a line $g$\ in the group plane if $P \mid g$. Two lines 
$g,h$ are said to be parallel if $g,h \nmid P,c$, for all $P,c$,
in other words, if they have neither a common perpendicular line nor a 
common point. Thus, if $P \neq Q$, then by A.1 and A.2, the
points $P$\ and $Q$\ in the group plane are joined by a unique line. If $%
P\nmid g$\ then A.6 says that there are precisely two lines through $P$\
parallel to $g.$

\begin{lemma} For each $\alpha \in \mathfrak{G}$, the mappings 
$\sigma_{a}:$\ $g\longmapsto g^{\alpha }\equiv \alpha g\alpha^{-1}$ and 
$\sigma_{\alpha }:P\longmapsto P^{\alpha }\equiv \alpha P\alpha^{-1}$ 
are one-to-one mappings of the set of lines and the set of points, each 
onto itself in the group plane.
\end{lemma}

\begin{proof} Let $\alpha \in \mathfrak{G},$\ and consider the mapping 
$\gamma \longmapsto \gamma ^{\alpha }\equiv \alpha \gamma \alpha^{-1}$ of 
$\mathfrak{G}$\ onto itself. It is easily seen that this mapping is
bijective. Since $\mathcal{G}$\ is an invariant system ($a^{b}\in \mathcal{G}
$\ for every $a\in \mathcal{G}, b \in \mathfrak{G}$) $\mathcal{G}$\ will 
be mapped onto
itself, and if $P$\ is a point, so that $P=gh$\ with $g|h,$\ then 
$P^{\alpha}=g^{\alpha }h^{\alpha }$\ and $g^{\alpha }|h^{\alpha }$, so that 
$P^{\alpha }$ is also a point. Thus, $g\longmapsto g^{\alpha }$, 
$P\longmapsto P^{\alpha }$\ are one-to-one mappings of the set of lines and
the set of points, each onto itself in the group plane.
\end{proof}

\begin{definition} A one-to-one mapping $\sigma $ of the set of points
and the set of lines each onto itself is called an \emph{orthogonal
collineation} if it preserves incidence and orthogonality.
\end{definition}

\nind Since the ``$|$'' relation is preserved under the above
mappings, orthogonal collineations also preserve incidence and
orthogonality as defined above.

\begin{cor} The mappings $\sigma _{\alpha }:g\longmapsto
g^{\alpha }$\ and $\sigma _{\alpha }:P\longmapsto P^{\alpha }$\ are
orthogonal collineations of the group plane and are called motions of the
group plane induced by $\alpha .$
\end{cor}

\nind In particular, if $\alpha $\ is a line $a$, we have a reflection
about the line $a$\ in the group plane, and if $\alpha $\ is a point
$A,$\ we have a point reflection about $A$\ in the group plane.

     If to every $\alpha \in \mathfrak{G}$\ one assigns the motion of the group
plane induced by $\alpha ,$\ one obtains a homomorphism of $\mathfrak{G}$\
onto the group of motions of the group plane. Bachmann shows in \cite{Ba} that
this homomorphism is in fact an isomorphism, so that points and lines in the
group plane may be identified with their respective reflections. Thus, 
$\mathfrak{G}$\ is seen to be the group of orthogonal collineations of 
$\mathfrak{G}$\ generated by $\mathcal{G}$.

\begin{definition} Planes which are representable as an isomorphic
image, with respect to incidence and orthogonality, of the group plane of a
group of motions $(\mathfrak{G},\mathcal{G}),$\ are called metric planes.
\end{definition}

     In \cite{BBPW}, BPWB show how one can embed a metric plane into a
projective--metric plane by constructing an ideal plane using pencils of
lines. We shall outline how this is done.

\begin{definition} Three lines are said to lie in a pencil if their
product is a line; {\it i.e.}, $a,b,c$\ lie in a pencil if 
\begin{equation}
abc=d \in \mathcal{G} \quad . \label{pencil}
\end{equation}
\end{definition}
\begin{definition} Given two lines $a,b$\ with $a\neq b,$\ the set
of lines $c$ satisfying equation (\ref{pencil}) is called a pencil of
lines and is denoted by $G(ab)$, since it depends only on the product
$ab.$
\end{definition}

     Note that the relation (\ref{pencil}) is symmetric, {\it i.e.} it
is independent of the order in which the three lines are taken: since
$cba=(abc)^{-1}$\ is a line, the invariance of $\mathcal{G}$\ implies
that $cab=(abc)^{c}$\ is a line and that every motion of the group
plane takes triples of lines lying in a pencil into triples in a
pencil. The invariance of $\mathcal{G}$\ also shows that
(\ref{pencil}) holds whenever at least two of the three lines
coincide.

     Using axioms implied by A.1 -- A.6, BPWB then show that there are
three distinct classes of pencils.

   (1) If $a,b | V$ then $G(ab)=\{c:c | V\}$. In this case, $G(ab)$ is
called a pencil of lines with center $V$ and is denoted by $G(V)$.

   (2) If $a,b$\ $|c$\ then $G(ab)=\{d:d | c\}$. In this case, $G(ab)$ is
called a pencil of lines with axis $c$ and is denoted by $G(c)$.

   (3) By A.6, there exist parallel lines $a,b$. Thus, in this case 
$G(ab)=\{c:c\parallel a,b$\ where $a\parallel b\},$\ which we denote by 
$G(ab)_{\infty }$. 

     An ideal projective plane $\Pi$ is constructed in the
following manner. An ideal point is any pencil of lines $G(ab)$\ of
the metric plane. The pencils $G(P)$\ correspond in a one-to-one way
to the points of the metric plane. An ideal line is a certain set of
ideal points.  There are three types:

   (1) A proper ideal line $g(a)$ is the set of ideal points which have in
common a line $a$ of the metric plane.

   (2) The set of pencils $G(x)$\ with $x$\ $|P$\ for a fixed point $P$ of
the metric plane, which we denote by $g(P)$.

   (3) Sets of ideal points which can be transformed by a 
halfrotation\footnote{see p. 161 in \cite{BBPW}} about
a fixed point $P$ of the metric plane into a proper ideal line; these we
denote by $g(ab)_{\infty }$.

     The polarity is defined by mapping $G(C)\longmapsto g(C)$ and 
$g(C)\longmapsto G(C);$\ $G(ab)_{\infty }\longmapsto g(ab)_{\infty }$\ and $%
g(ab)_{\infty }\longmapsto G(ab)_{\infty };$ and $G(c)\longmapsto g(c)$ and $%
g(c)\longmapsto G(c).$\ In \cite{Ba}, Bachmann shows that the resulting ideal
plane is a hyperbolic projective plane in which the theorem of Pappus and
the Fano axiom both hold, {\it i.e.} a hyperbolic projective--metric plane.

     In this model, the ideal points of the form $G(P)$\ are the
interior points of the hyperbolic projective--metric plane; thus the
points of the metric plane correspond in a one--to--one manner with the
interior points of the hyperbolic projective--metric plane. The ideal
points $G(x),$\ for $x\in \mathcal{G}$\ are the exterior points of the
hyperbolic projective--metric plane.

\begin{prop} Each $x\in \mathcal{G}$\ corresponds in a one-to-one
manner with the exterior points of the hyperbolic projective--metric plane.
\end{prop}

\begin{proof} Since each line $d$\ of the metric plane is incident with at
least three points (Theorem 5 in \cite{BBPW}) and a point is of the form $ab$
with $a | b$, it follows that each $x\in\mathcal{G}$ is the axis of a 
pencil. From the uniqueness of perpendiculars (Theorem 4 \cite{BBPW}), 
each $x\in \mathcal{G}$\ corresponds in a one-to-one manner with the 
pencils $G(x)$. Hence, each $x\in \mathcal{G}$ corresponds in a one-to-one 
manner with the exterior points of the hyperbolic projective--metric plane.
\end{proof}

     Thus, one may view the axioms as referring to the interior and exterior
points of a hyperbolic projective--metric plane. The ideal points of the
form $G(ab)_{\infty }$, where $a\parallel b$, are the points on the
absolute, {\it i.e.}, the points at infinity in the hyperbolic
projective--metric plane.

     We turn to the ideal lines. A proper ideal line $g(a)$\ is a set of
ideal points which have in common a line $a$\ of the metric plane.

\begin{prop} A proper ideal line $g(a)$\ is a secant line of the
form \newline
$g(a)=\{G(P),G(x),G(bc)_{\infty }:x,P|a$\ and $abc\in \mathcal{G}$\ \
where $b\parallel c\}.$
\end{prop}

\begin{proof} By Theorem 23 of \cite{BBPW}, any two pencils of lines of the
metric plane have at most one line in common. By A.6, each line belongs to at
least two pencils of parallels and by Theorem 13 of \cite{BBPW} and A.6 
again, each line $g\in \mathcal{G}$\ belongs to precisely two such pencils. 
Thus, a proper ideal line contains two points on the absolute, interior 
points, and exterior points. Hence, a proper ideal line is a secant line. 
A secant line is the set \newline
$g(c)=\{G(P),G(x),G(ab)_{\infty }:x,P\ |c$\ and $abc\in 
\mathcal{G}$\ where $a\parallel b\}$.
\end{proof}

\begin{cor} The ideal line which consists of pencils $G(x)$\
with $x$\ $|P$\ \ for a fixed point $P$\ of the metric plane consists only of
exterior points, {\it i.e.}, it is an exterior line. Therefore, 
$g(P)=\{G(x):x|P\}.$
\end{cor}

     The final type of ideal line is a tangent line. It contains only one 
point $G(ab)_{\infty }$\ on the absolute. Denoting this line by 
$g(ab)_{\infty }$, we have 
$g(ab)_{\infty }=\{G(ab)_{\infty }\}\cup \{G(x):x\in \mathcal{G}$ and $%
abx\in \mathcal{G}\}$\ where $a\parallel b.$\ Recalling that each $x\in 
\mathcal{G}$\ corresponds to an exterior point in the hyperbolic
projective--metric plane, we see that a tangent line consists of one point on
the absolute and every other point is an exterior point.

     We also note that under the above identifications, each secant line $%
g(c)$\ corresponds to a unique ``exterior point'' $G(c),$\ $G(c)\notin g(c)$,
since one only considers those $x,P$\ $|c$\ such that 
$xc\neq 1_{\mathfrak{G}}$ and $Pc\neq 1_{\mathfrak{G}}$. Each exterior line
corresponds to a unique interior point $P$ and each tangent line
corresponds to a unique point on the absolute.

\begin{prop} The map $\Phi$ given by

(i) $\ \ \Phi (G(c))=g(c),$\ $\Phi (g(c))=G(c)$

(ii) $\ \Phi (G(P))=g(P),$\ $\Phi (g(P))=G(P)$

(iii) $\Phi (G(ab)_{\infty })=g(ab)_{\infty },$\ $\Phi (g(ab)_{\infty
})=G(ab)_{\infty }$

is a polarity.
\end{prop}

\begin{proof} Let $\mathcal{P}$ be the set of all points of $\Pi$ and 
$\mathcal{L}$ the set of all lines of $\Pi$. From the remarks above
it follows that $\Phi$ is a well-defined one-to-one point--to--line mapping
of $\mathcal{P}$ onto $\mathcal{L}$ and a well-defined one-to-one
line--to--point mapping of $\mathcal{L}$ onto $\mathcal{P}$. 

     Next, it is shown that $\Phi$ is a correlation and for this it 
suffices to show that $\Phi$ preserves incidence. Let 
$g(c)=\{G(P),G(x),G(ab)_{\infty }:x,P|c$ and $abc\in \mathcal{G}$
where $a\parallel b\}$ be a secant line. Let 
$G(A),G(B),G(d),G(ef)_{\infty }\in g(c),$ where 
$G(ef)_{\infty }=\{x\in \mathcal{G}:xef\in \mathcal{G}$ and 
$e\parallel f\}.$ Then $A,B,d|c$ and $cab\in \mathcal{G}.$ Now 
$\Phi(G(A))=g(A)=\{G(x):x|A\},$ $\Phi(G(B))=g(B)=\{G(x):x|B\},$ 
$\Phi(G(d))=g(d)$, and $\Phi(G(ef)_{\infty })= g(ef)_{\infty }=\{G(ef)_{\infty
}\}\cup \{G(x):efx\in \mathcal{G}\}.$ Thus, it follows that 
$\Phi(g(c))=G(c)\in g(A)\cap g(B)\cap g(d)\cap g(ef)_{\infty },$ so that 
$\Phi(g(c))\in \Phi(G(A)),\Phi(G(B)),\Phi(G(d)),\Phi(G(ef)_{\infty })$ 
and $\Phi$ preserves incidence on a secant line.

     Consider an exterior line $g(P)=\{G(x):x|P\}$, and let 
$G(a),G(b)\in g(P)$. Then $a,b|P$ and it follows that 
$G(P)\in g(a)\cap g(b)$, {\it i.e.} 
$\Phi(g(P))\in \Phi(G(a))\cap \Phi(G(b))$ and $\Phi$ preserves incidence
on an exterior line.

     Finally, let $g(ab)_{\infty }=\{G(ab)_{\infty }\}\cup \{G(x):abx\in 
\mathcal{G}$ where $a\parallel b\}$ be a tangent line. Clearly, since 
$\Phi(G(ab)_{\infty })=g(ab)_{\infty }$, one has 
$G(ab)_{\infty }\in g(ab)_{\infty }$. 
Now suppose that $G(d)\in g(ab)_{\infty }.$ Then $abd\in \mathcal{G}$ and
$\Phi(G(d))=g(d)=\{G(A),G(x),G(ef)_{\infty}:A,x|d$ and $def\in \mathcal{G}$ 
where $e\parallel f\}.$ Thus, $G(d)\in G(ab)_{\infty }\cap G(ef)_{\infty}$ 
and $G(ef)_{\infty }\in g(d)$, which implies that $G(d)\in g(ef)_{\infty}$ 
and $\Phi(g(ef)_{\infty })\in \Phi(G(d))$. Hence, $\Phi$
preserves incidence and is a correlation.
\end{proof}

     It follows from the above observations that $\Phi$ transforms
the points $G(Y)$ on a line $g(b)$ into the lines $\Phi(G(Y))$
through the point $\Phi(g(b))$. Thus, $\Phi$ is a projective
correlation. Since $\Phi^{2}$ is the identity map, then $\Phi$ is a
polarity. Moreover, since $\Phi(g(ab)_{\infty})=G(ab)_{\infty }$
with $G(ab)_{\infty}\in g(ab)_{\infty},$ then $\Phi$ is a
hyperbolic polarity.

\begin{prop} The definition of orthogonality given by the polarity
coincides with and is induced by the definition of orthogonality in 
the group plane.
\end{prop}

\begin{proof} Declaring a perpendicularity with respect to the
polarity defined above, one has, on the one hand, $g(c) \perp g(a)$ 
if and only if $\Phi(g(c))= G(c) \in g(a)$ and, on the other, 
$\Phi(g(a))= G(a)\in g(c)$ if and only if $a | c$. Similarly, one has
$g(c) \perp g(P)$ if and only if $\Phi(g(c))= G(c) \in g(P)$ if and
only if $c | P$. In addition, $g(P) \perp g(Q)$ if and only if 
$\Phi(g(P))=G(P)\in g(Q)$ if and only if $P | Q$. In fact, this is
excluded by Theorem 2.1(c) in \cite{BBPW}, in conformity with hyperbolic 
geometry, since two interior points cannot be conjugate under the 
hyperbolic polarity.

     Further, one sees that $G(C) \perp G(p)$ if and only if 
$G(C)\in \Phi(G(p))=g(p)$ and $G(p)\in \Phi(G(C))=g(C)$ if and only if 
$P | c$. Finally, one also has $G(c) \perp G(x)$ if and only if 
$G(c)\in \Phi(G(x))=g(x)$ and $G(x)\in g(c)=\Phi(G(c))$ if and only if
$x | c$.
\end{proof}

     We also see that if instead of interpreting our original
generators as ordinary lines in a hyperbolic plane, we interpret them
as exterior points, then we can construct a hyperbolic
projective--metric plane in which the theorem of Pappus and Fano's
axiom both hold and which is generated by the exterior points of the
hyperbolic projective--metric plane.

     With the identifications and the geometric objects defined above,
we show in the next subsection that the motions of the hyperbolic
projective--metric plane above can be generated by reflections about
exterior points, {\it i.e.} any transformation in the hyperbolic plane
which leaves the absolute invariant can be generated by a suitable
product of reflections about exterior points.

\subsection{Reflections About Exterior Points}

     In \cite{Cox3} Coxeter showed that any congruent
transformation of the hyperbolic plane is a collineation which preserves the
absolute and that any such transformation is a product of reflections about
ordinary lines in the hyperbolic plane, where a line reflection about a line 
$m$\ is a harmonic homology with center $M$\ and axis $m$, where $M$\ and $m$
are a pole--polar pair and $M$\ is an exterior point. A point reflection is
defined similarly: a harmonic homology with center $M$\ and axis $m,$\ where 
$M$\ and $m$\ are a pole-polar pair, $M$\ is an interior point, and $m$\ is
an exterior line. Note that in both cases, $M$\ and $m$\ are nonincident.
We recall a series of definitions for the convenience of the reader.

\begin{definition} A collineation is a one-to-one map of the set of
points onto the set of points and a one-to-one map of the set of lines onto
the set of lines that preserves the incidence relation.
\end{definition}

\begin{definition} A perspective collineation is a collineation which
leaves a line pointwise fixed --- called its axis --- and a point 
line-wise fixed --- called its center.
\end{definition}

\begin{definition} A homology is a perspective collineation with
center a point $B$\ and axis a line $b$\ where $B$\ is not incident with $b.$
\end{definition}

\begin{definition} A harmonic homology with center $B$\ and axis $b,$\
where $B$\ is not incident with $b$, is a homology which relates each point $%
A$\ in the plane to its harmonic conjugate with respect to the two points $B$%
\ and $(b,[A,B]),$\ where $[A,B]$\ is the line joining $A$\ and $B$\ and $%
(b,[A,B])$\ is the point of intersection of $b$\ and $[A,B].$
\end{definition}

\begin{definition} A complete quadrangle is a figure consisting of
four points (the vertices), no three of which are collinear, and of the six
lines joining pairs of these points. If $l$\ is one of these lines, called a
side, then it lies on two of the vertices, and the line joining the other
two vertices is called the opposite side to $l$. The intersection of two
opposite sides is called a diagonal point.
\end{definition}

\begin{definition} A point $D$\ is the harmonic conjugate of a point $%
C $\ with respect to points $A$\ and $B$\ if $A$\ and $B$\ are two vertices
of a complete quadrangle, $C$\ is the diagonal point on the line joining $A$%
\ and $B,$\ and $D$\ is the point where the line joining the other two
diagonal points cuts $[A,B].$\ One denotes this relationship by $H(AB,CD).$
\end{definition}

     In keeping with the notation employed at the end of $\mathcal{x}\, 2.1,$\ 
let $G(b)$\ be an exterior point and $g(b)$\ its pole.

\begin{lemma} The map $\Psi _{b}:\left\{ 
\begin{array}{ccccccc}
G(A) & \longmapsto  & G(A)^{b} & and & G(d) & \longmapsto  & G(d)^{b} \\ 
g(A) & \longmapsto  & g(A)^{b} &  & g(d) & \longmapsto  & g(d)^{b} \\ 
G(cd)_{\infty } & \longmapsto  & G(cd)_{\infty }^{b} &  & g(cd)_{\infty } & 
\longmapsto  & g(cd)_{\infty }^{b}%
\end{array}%
\right\} $\ is a collineation.
\end{lemma}

\begin{proof} This follows from the earlier observation that the motions
of the group plane map pencils onto pencils preserving the ``$|"$\ relation.
\end{proof}

\begin{lemma} $\Psi _{b}$\ is a perspective collineation and, hence, a
homology.
\end{lemma}

\begin{proof} Recall that $g(b)=\{G(A),G(x),G(cd)_{\infty }:x,A$ $|b$
and where $b$ lies in the pencil $G(cd)_{\infty }\}$. For any $G(A)$
and $G(x)$\ in $g(b)$ one has $A^{b}=A$ and $x^{b}=x$, since $A,x$\ $|b$
and if $G(A),G(x)\notin g(b)$\ then 
$A,x \nmid b$ and $A^{b}\neq A,$\ $x^{b}\neq x,$ and 
$A^{b},x^{b}\nmid b$. Thus, $G(A)^{b},G(x)^{b}\notin g(b)$.

   Recall also that $G(cd)_{\infty} = \{f \mid fcd\in \mathcal{G},$
where $c$ and $d$ have neither a common point nor a common perpendicular $\}$. 
Now $g(b)$ is a secant line, so that it contains two such distinct points, 
$G(mn)_{\infty}$ and $G(pq)_{\infty }$, say, on the absolute. Since the
motions of the group plane map pencils onto pencils preserving the ``$|$''
relation it follows that if $c,d\in G(mn)_{\infty}$ then 
$c^{b},d^{b}\in G(mn)_{\infty }$ and hence, 
$G(mn)_{\infty}^{b}=G(mn)_{\infty }$ and $G(pq)_{\infty }^{b}=G(pq)_{\infty}$.
Moreover, if $G(rs)_{\infty }\notin g(b)$, then it follows that 
$G(rs)_{\infty }^{b}\notin g(b)$. Thus, $\Psi_{b}$ leaves $g(b)$
pointwise invariant.

     Let $g(d), g(Q)$, and $g(rs)_{\infty }$\ be a secant line,
exterior line, and tangent line, respectively, containing $G(b).$ For 
$G(e)\in g(d)$ one has $e$\ $|d$\ and $e^{b}\mid d^{b}=d$, since $b\mid d$,
thus $G(e)^{b}\in g(d)$. For $G(A)\in g(d)$, $A^{b}$\ $|d^{b}=d$, so 
$G(A)^{b}\in g(d)$. Similarly, it follows that if $G(ef)_{\infty }\in g(d)$
then $G(ef)_{\infty }^{b}\in g(d)$, and $g(d)^{b}=g(d)$. One easily
sees that $g(Q)^{b}=g(Q)$ and $g(rs)_{\infty }^{b}=g(rs)_{\infty }$.
Thus, $\Psi _{b}$ leaves every line through $b$ invariant and $\Psi_{b}$
is a perspective collineation for each $b\in \mathcal{G}$.
\end{proof}

\begin{prop} $\Psi_{b}$\ is a harmonic homology.
\end{prop}

\begin{proof} Since $A^{b}$\ is again a point in the original group plane
and since $d^{b}$\ is again a line in the original group plane, 
it follows that, for each 
$b\in\mathcal{G}$, $\Psi _{b}$\ maps interior points to interior points,
exterior points to exterior points, points on the absolute to points
on the absolute, secant lines to secant lines, exterior lines to
exterior lines, and tangent lines to tangent lines.  Moreover, since
$(\xi^{b})^{b} = \xi$ for any $\xi \in \mathfrak{G}$, $\Psi_{b}$\ is
involutory for each $b\in \mathcal{G}$. But in a projective plane in
which the theorem of Pappus holds, the only collineations which are
involutory are harmonic homologies \cite{BK}. Thus $\Psi_{b}$ is a
harmonic homology for each $b\in \mathcal{G}$.
\end{proof}

\begin{prop} Point reflections about interior points are generated
by reflections about exterior points.
\end{prop}

\begin{proof} Arguing in a similar manner, one sees that for each 
interior point $G(A)$, $\Psi_{A}$ is a harmonic homology with center 
$G(A)$ and axis $g(A)$, where $g(A)$ is the polar of $G(A)$,
$G(A)\notin g(A)$, and where $\Psi_{A}$ is defined analogously to 
$\Psi_{b}$. Thus, each $\Psi_{A}$ is a point reflection, and since 
$A$ is the product of two exterior points, one sees that point reflections 
about interior points are generated by reflections about exterior points.
\end{proof}

\begin{prop} The reflection of an interior point about a secant
line coincides with the reflection of the same interior point about 
an exterior point. Moreover, since any motion of the hyperbolic plane 
is a product of line reflections about secant lines, any motion of the 
hyperbolic plane is generated by reflections about exterior points.
\end{prop}

\begin{proof} Consider a line reflection in the hyperbolic plane, {\it i.e.}
the harmonic homology with axis $g(b)$ and center $G(b)$. Let $G(A)$
be an interior point and $g(d)$ a line through $G(A)$ meeting $g(b)$.
Since $G(b)\in g(d)$, one has $b | d$ and $g(d)$ is orthogonal to $g(b)$.
Let $G(E)$ be the point where $g(b)$ meets $g(d)$. Since $G(E)\in g(b)$,
then $E | b$ and $Eb=f$ for some $f\in \mathcal{G}$. It follows
that the reflection of $G(A)$ about $g(b)$ is the same as the reflection
of $G(A)$ about $G(E)$. Since $b | d$ and $E|d$, then $bd=C$ and
one has $E,C | b,d$ with $b\neq d$. Thus, by A.2, $E=C=bd$. Hence,
$A^{E}=A^{db}$ and $A|d$ as $G(A)\in g(d) = A^{b}$.
\end{proof}

     Since the motions of the projective--metric plane are precisely those
collineations which leave the absolute invariant, we have the 
following result.

\begin{theorem} Reflections of exterior points about exterior points
and about exterior lines are also motions of the projective--metric plane.
Hence, the $\Psi_{b}$'s for $b\in \mathcal{G}$ acting on exterior points
and exterior lines are motions of the hyperbolic projective--metric plane.
\end{theorem} 

     We also point out that the proof that each $\Psi_{b}$ is an involutory
homology also shows that the Fano axiom holds, since in a projective plane
in which the Fano axiom does not hold no homology can be an involution 
\cite{BL}.

\subsection{Embedding a Hyperbolic Projective--Metric Plane Into a Projective
3-Space}

     We embed our hyperbolic projective--metric plane into a
three-dimensional projective space, finally obtaining an affine space
whose plane at infinity is isomorphic to our original
projective--metric plane. Any projective plane $\Pi$ in which the
theorem of Pappus holds can be represented as the projective
coordinate plane over a field $\mathcal{K}$. (The theorem of Pappus
guarantees the commutativity of $\mathcal{K}$.) Then by considering
quadruples of elements of $\mathcal{K}$, one can define a projective
space $P_{3}(\mathcal{K})$ in which the coordinate plane corresponding
to $\Pi$ is included. Since the Fano axiom holds, the corresponding
coordinate field $\mathcal{K}$ is not of characteristic 2
\cite{BL}. In fact, A.6 entails that $\mathcal{K}$ is a Euclidean
field. By singling out the coordinate plane corresponding to $\Pi$
as the plane at infinity, one obtains an affine space whose plane at
infinity is a hyperbolic projective--metric plane, {\it i.e.}
three-dimensional Minkowski space.

     To say that a plane $\Pi$ is a projective coordinate plane over a field 
$\mathcal{K}$ means that each point of $\Pi$ is a triple of numbers 
$(x_{0},x_{1},x_{2})$, not all zero, together with all multiples 
$(\lambda x_{0},\lambda x_{1},\lambda x_{2})$, $\lambda \neq 0$. Similarly,
each line of $\Pi$ is a triple of numbers $[u_{0},u_{1},u_{2}]$, not all 
zero, together with all multiples 
$[\lambda u_{0},\lambda u_{1},\lambda u_{2}],$\ $\lambda \neq 0$. 
In P$_{3}(\mathcal{K})$ all the quadruples of
numbers with the last entry zero correspond to $\Pi$. One can now obtain an
affine space $\mathcal{A}$\ by defining the points of $\mathcal{A}$ to be
those of P$_{3}(\mathcal{K})-\Pi $, {\it i.e.} those points whose last entry
is nonzero; a line $l$\ of $\mathcal{A}$\ to be a line $l^{\prime }$ in 
$P_{3}(\mathcal{K})-\Pi $ minus the intersection point of the line 
$l^{\prime}$ with $\Pi$; and by defining a point $P$ in $\mathcal{A}$ to be
incident with a line $l$ of $\mathcal{A}$ if and only if $P$ is
incident with the corresponding $l^{\prime}$. Planes of $\mathcal{A}$ are
obtained in a similar way \cite{Cox4}.

   Thus, each point in $\Pi$ represents the set of all lines in $\mathcal{A}$
parallel to a given line, where lines and planes are said to be parallel
if their first three coordinates are the same, and each line in $\Pi$
represents the set of all planes parallel to a given plane. Since parallel
objects can be considered to intersect at infinity, we call $\Pi$ the
plane at infinity.

\subsubsection{Exterior Point Reflections Generate Motions in an Affine Space}

      In \cite{Cox1}, Coxeter
shows that three-dimensional Minkowski space is an affine space whose plane
at infinity is a hyperbolic projective--metric plane. He also classifies the
lines and planes of the affine space according to their sections by the
plane at infinity:

\medskip

\underline{Line or Plane}\ \ \ \ \ \ \ \ \ \ \ \ \ \ \ \ \ \ \ \ \ \ \ \ \ \
\ \ \ \ \ \ \ \ \ \ \ \ \ \ \ \ \ \ \ \ \ \ \ \ \ \ \ \ \ \ \underline{%
Section at Infinity}

Timelike line \ \ \ \ \ \ \ \ \ \ \ \ \ \ \ \ \ \ \ \ \ \ \ \ \ \ \ \ \ \ \
\ \ \ \ \ \ \ \ \ \ \ \ \ \ \ \ \ \ \ \ \ \ \ \ Interior point

Lightlike line \ \ \ \ \ \ \ \ \ \ \ \ \ \ \ \ \ \ \ \ \ \ \ \ \ \ \ \ \ \ \
\ \ \ \ \ \ \ \ \ \ \ \ \ \ \ \ \ \ \ \ \ \ \ \ Point on the absolute

Spacelike line \ \ \ \ \ \ \ \ \ \ \ \ \ \ \ \ \ \ \ \ \ \ \ \ \ \ \ \ \ \ \
\ \ \ \ \ \ \ \ \ \ \ \ \ \ \ \ \ \ \ \ \ \ \ Exterior point

Characteristic plane \ \ \ \ \ \ \ \ \ \ \ \ \ \ \ \ \ \ \ \ \ \ \ \ \ \ \ \
\ \ \ \ \ \ \ \ \ \ \ \ \ \ \ \ \ Tangent line

Minkowski plane \ \ \ \ \ \ \ \ \ \ \ \ \ \ \ \ \ \ \ \ \ \ \ \ \ \ \ \ \ \
\ \ \ \ \ \ \ \ \ \ \ \ \ \ \ \ \ \ \ \ Secant line

Spacelike plane \ \ \ \ \ \ \ \ \ \ \ \ \ \ \ \ \ \ \ \ \ \ \ \ \ \ \ \ \ \
\ \ \ \ \ \ \ \ \ \ \ \ \ \ \ \ \ \ \ \ \ \ Exterior line

\medskip

\nind He shows that if one starts with an affine space and introduces a
hyperbolic polarity in the plane at infinity of the affine space, then the
polarity induces a Minkowskian metric in the whole space. With this
hyperbolic polarity one considers as perpendicular any line and plane or any
plane and plane whose elements at infinity correspond under this polarity.
Two lines are said to be perpendicular if they intersect and their elements
at infinity correspond under the polarity.

     The proof that the group of motions of three-dimensional Minkowski
space is generated by the ``reflections about spacelike lines'' defined
above is the final step of our argumentation in this section. 

\begin{theorem} \label{main}
Exterior point reflections generate any motion in the affine space. 
Moreover, since exterior points correspond to spacelike lines,
any motion in Minkowski 3-space is generated by reflections about
spacelike lines.
\end{theorem}

\begin{proof} Since any motion in Minkowski space can be generated by a
suitable product of plane reflections, it suffices to show that reflections
about exterior points generate plane reflections.

     Let $\alpha$ be any Minkowski plane or spacelike plane (note that
reflections about characteristic planes and lightlike lines do not exist
since they are self-perpendicular or see \cite{BK}). Let $P$ be any point in
Minkowski space. Let $l$ be the line through $P$ parallel to $\alpha$.
Let $\alpha_{\infty }$ denote the section of $\alpha$ at infinity.
Applying the polarity to $\alpha_{\infty }$, one obtains a point 
$g_{\infty} \perp \alpha_{\infty }$. Let $g$ be a line through $P$ whose 
section at infinity is $g_{\infty}$, so that $g$ is a line through $P$
orthogonal to $\alpha$. Since each line in the plane at infinity contains
at least 3 points, there exists a line $l$ in $\alpha$ which is
orthogonal to $g$ as $g_{\infty} \perp \alpha_{\infty}$. Now let 
$m$ be a line through $P$ not in $\alpha$ which intersects $l$. It
follows that the reflection of $P$ about $\alpha$ is the same as
reflecting $m$ about $l$ and taking the intersection of the image of $m$
under the reflection with $g$. By the construction of the affine space and
the definition of orthogonality in the affine space, it follows that $l$ and 
$m$ must act as their sections at infinity act. Since any point reflection
in the hyperbolic projective--metric plane can be generated by reflections
about exterior points, it follows that the reflection of $P$ about $\alpha$
is generated by reflections of $P$ about spacelike lines.
\end{proof}

     From $(\mathfrak{G},\Gs)$ we have therefore constructed a model
of three-dimensional Minkowski space in which each element of $\Gs$ is
identified as a spacelike line (and every spacelike line in the
Minkowski space is such an element) and on which each element of $\Gs$
acts adjointly as the reflection about the spacelike line. Such
reflections generate the proper Poincar\'e group $\Ps_+$ on
three-dimensional Minkowski space. The group $\mathfrak{G}$ is
therefore isomorphic to $\Ps_+$, and the adjoint action of the identity
component of $\mathfrak{G}$ upon $\Gs$ is transitive.

\section{From States and Observables to Space--Time}

     An operationally motivated and mathematically powerful approach
to quantum field theory is algebraic quantum field theory (AQFT) (cf.
\cite{Haag}). The initial data in AQFT are a collection $\{\As(\Os)\}$
of unital $C^*$-algebras indexed by a suitable set of open subregions
$\Os$ of the space--time of interest, with $\As(\Os)$ understood as
being generated by all the observables measurable in the spacetime
region $\Os$, and a state $\omega$ on these algebras, understood as
representing the preparation of the quantum system under
investigation. For the reasons mentioned in the Introduction, we shall
replace the index set of subregions of a specified space--time with
some abstract set $I$, which for our purposes may be viewed as
indexing possible laboratories.  Hence, $\As_i$ is interpreted as the
algebra generated by all observables measurable in the ``laboratory'' 
$i \in I$. It is understood that the description of the laboratory would
include not only ``spatial'' but also ``temporal'' specifications. 
These specifications would be made with respect to suitable measuring
devices, which themselves do not presuppose a particular space--time.
There may be some structure on the index set $I$. For example,
it makes sense to represent the fact that laboratory $i$ is contained
in the laboratory $j$ with $i < j$. Then one would certainly have
the relation $\As_i \subset \As_j$, {\it i.e.} $\As_i$ is a subalgebra
of $\As_j$.\footnote{This property of the net $\inet$ is called isotony 
in the AQFT literature.} 
Hence, if $(I,\leq)$ is a partially ordered set, then one may expect
that the property of isotony holds. We would therefore be working with two 
partially ordered sets, $(I,\leq)$ and $(\{ \As_i \}_{i \in I}, \subseteq)$, 
and we require that the assignment $i \mapsto \As_i$ be an order-preserving 
bijection ({\it i.e.} it is an isomorphism in the structure class of partially 
ordered sets). Any such assignment which is not an isomorphism in 
this sense would involve some kind of redundancy in the description. To 
different laboratories should correspond different algebras.\footnote{This
truism need not hold in certain space--times such as anti-de Sitter 
space--time \cite{BFS3,BFS4}, where there exist closed timelike curves.}

     If $\inet$ is a net, then the inductive limit $\As$ of $\inet$ exists
and may be used as a reference algebra. However, even if $\inet$ is not a net, 
it is still possible \cite{Fr} to naturally embed the algebras 
$\As_i$ in a $C^*$-algebra $\As$ in such a way that the inclusion relations 
are preserved. In the following we need not distinguish these two
cases and refer, somewhat loosely, to any collection $\{ \As_i \}_{i \in I}$ 
of algebras, as specified, as a net. Any  state on $\As$ restricts to
a state on $\As_i$, for each $i \in I$. For that reason, we shall speak of a 
state on $\As$ as being a state on the net $\inet$. 

     Given a state $\omega$ on the algebra $\As$, one can consider the 
corresponding GNS representation $(\Hs_{\omega},\pi_{\omega},\Omega)$ and the 
von Neumann algebras $\Rs_i \equiv \pi_{\omega}(\As_i)''$, $i \in I$. We
shall assume that the representation space $\Hs_{\omega}$ is separable. We 
extend the assumption of nonredundancy of indexing to the net 
$\{ \Rs_i \}_{i \in I}$, {\it i.e.} we assume that also the map 
$i \mapsto \Rs_i$ is an order-preserving bijection.\footnote{This is 
automatically the case if the algebras $\As_i$ are $W^*$-algebras and
$\omega$ induces a faithful representation of 
$\cup_{i \in I} \As_i$.} If the GNS vector $\Omega$ is cyclic and 
separating for each algebra $\Rs_i$, $i \in I$, then from the modular theory 
of Tomita-Takesaki \cite{Tak,BR}, we are presented with a collection 
$\{J_i\}_{i\in I}$ of modular involutions 
(and a collection $\{\Delta_i\}_{i\in I}$ of modular 
operators), directly derivable from the state and the algebras. This 
collection $\{J_i\}_{i\in I}$ of operators on $\Hs_{\omega}$ generates a group 
$\Js$. Note that $J\Omega = \Omega$ for all $J \in \Js$. 
In the following we shall denote the adjoint action of $J_i$ upon the 
elements of the net $\irnet$ by $\ad J_i$, {\it i.e.} 
$\ad J_i (\Rs_j) \equiv J_i \Rs_j J_i = \{ J_i A J_i : A \in \Rs_j \}$. 
Note that if $\Rs_1 \subset \Rs_2$, then one necessarily has 
$\ad J_i (\Rs_1) \subset \ad J_i (\Rs_2)$, in other words the map $\ad J_i$ is
order-preserving. 

     The Condition of Geometric Modular Action (CGMA) was first
introduced in \cite{BS} and has received a great deal of development
since then --- see, {\it e.g.}, \cite{Bor0,BDFS,BFS1,BMS,Bor}. In the
present abstract setting, the CGMA is the condition that each map 
$\ad J_i$ leaves the set $\{ \Rs_i \}_{i \in I}$ invariant, 
{\it i.e.} $\ad J_i$ is a net
automorphism, for each $i \in I$. By the uniqueness of the modular
objects, it follows that $\{ J_i \}_{i \in I}$ is an invariant
generating set of involutions for the group $\Js$, as required for the
purposes of absolute geometry. Furthermore, we note that the CGMA implies 
that, for each $i \in I$, there exists an order-preserving bijection 
$\tau_i$ on $I$ such that $\ad J_i(\Rs_j) = \Rs_{\tau_i(j)}$ and 
$J_i J_j J_i = J_{\tau_i(j)}$, $i,j \in I$ \cite{BDFS}. The group generated 
by the involutions $\tau_i$, $i \in I$, is denoted by $\Ts$ and forms a
subgroup of the permutation group on the index set $I$. The set 
$\{ \tau_i \}_{i \in I}$ is also an invariant generating set of involutions
for the group $\Ts$. Thus, the pair $(\Ts,\{ \tau_i \}_{i \in I})$ also 
provides a candidate for an absolute geometric treatment. In fact, as 
shown in \cite{BDFS}, the group $\Js$ is a central extension of $\Ts$ by a 
subgroup $\Zs$ of the center of $\Js$. So, in general, one should possibly
consider the pair $(\Ts,\{ \tau_i \}_{i \in I})$ as the initial data
for the considerations of the previous section. However, in the case
we are examining, the center of $\Js$ turns out to be trivial, so that
$\Js$ and $\Ts$ are isomorphic. Hence, we shall avoid some technical
complications and impose the axioms A.1 -- A.6 on the pair
$(\Js, \{ J_i \}_{i \in I})$. And to avoid certain degeneracies, we
shall assume all algebras $\As_i$ to be nonabelian.

     For the convenience of the reader, we summarize our standing assumptions. 

\medskip

\nind{\bf Standing Assumptions} For the net $\inet$ of nonabelian 
$C^*$-algebras and the state $\omega$ on $\As$ we assume \par
   (i) $i \mapsto \Rs_i$ is an order-preserving bijection; \par
   (ii) $\Omega$ is cyclic and separating for each algebra $\Rs_i$, $i \in I$;
\par
   (iii) each $\ad J_i$ leaves the set $\irnet$ invariant.

\medskip

\nind Already these assumptions restrict significantly the class of
admissible groups $\Ts$ and $\Js$ \cite{BDFS}. In general, it may be
necessary to pass to a suitable subcollection of $\{ \Rs_i \}_{i \in I}$ 
in order for the Standing Assumptions to be satisfied \cite{BDFS} ---
see the final section for a brief discussion of this point. 
Note also that the Standing Assumptions imply
\begin{equation}
\Rs_{\tau_i(i)} = J_i \Rs_i J_i = \Rs_i{}' \quad , \label{ambiguity}
\end{equation}
for all $i \in I$. Hence, the surjective map $i \mapsto J_i$ is
two-to-one, since $J_i = J_{\tau_i(i)}$.

     Consider three-dimensional Minkowski space with the standard metric

$$g = \diag(1,-1,-1) \equiv 
\left( \begin{array}{ccc}
               1 & 0 & 0  \\ 
               0 & -1 & 0  \\
               0 & 0 & -1 
\end{array} \right) $$

\nind given in proper coordinates. The isometry group of this 
space is the Poincar\'e group $\Ps$ and the family $\Ws$ of wedges 
is obtained by applying the elements of $\Ps$ to a single wedge-shaped region 
of the form
$$W_R \equiv \{ x \in \RR^3 : x_1 > \vert x_0 \vert \} \quad , $$
{\it i.e.} $\Ws = \{ \lambda W_R : \lambda \in \Ps \}$, where
$\lambda W_R = \{ \lambda(x) : x \in W_R \}$. We remark that,
in fact, one has $\Ws = \{ \lambda W_R : \lambda \in \Pid \}$, where
$\Pid$ is the identity component of the Poincar\'e group. Each wedge
$W = \lambda W_R$ determines a spacelike line called the edge $E_W$ of 
the wedge: $E_W = \lambda E_R$, where
$$E_R = \{ x \in \RR^3 : x_1 = 0 = x_0 \} \quad . $$
Note that $W$ and its causal complement\footnote{The causal complement of a
set $S \subset \RR^3$ is the interior of the set of all points in $\RR^3$
which are spacelike separated from every point in $S$.} $W'$ share the 
same edge, {\it i.e.} $E_W = E_{W'}$. Moreover, the equality 
$E_{W_1} = E_{W_2}$ entails that either $W_1 = W_2$ or $W_1 = W_2{}'$. 
Conversely, each spacelike line $l \subset \RR^3$ determines a pair of 
wedges $W_1,W_2$ with $W_1 = W_2{}'$ --- in fact, the causal complement of
$l$ consists of two connected components, each a wedge, each the causal 
complement of the other, and each having $l$ as its edge.

     Now assume that $\omega$ and $\inet$ satisfy the Standing Assumptions, 
and that the pair $(\Js,\{ J_i \}_{i \in I})$ fulfills Axioms A.1 -- A.6 
in Section 2.1. With $\mathfrak{G} = \Js$ and $\Gs = \{ J_i \}_{i \in I}$,
the results of Section 2 entail that there exists a realization of
three-dimensional Minkowski space $\RR^3$ in which each $J_i$ corresponds
uniquely to a spacelike line $l_i$ and on which each $J_i$ acts 
adjointly as the reflection about $l_i$. Hence, to each $J_i$ corresponds a 
pair $W_i,W_i{}'$ of complementary wedges whose common edge is $l_i$
(recall that in this construction $l_i$ is, in fact, equal to $J_i$). 
One of these spacelike lines is the set 
$l_{i_0} \equiv E_R = \{ (0,0,x) : x \in \RR \}$. Define 
$\chi(i_0) \equiv W_R$. 
From the results in Section 2, the adjoint action of the identity
component (isomorphic to $\Pid$) of the group $\Js$ 
upon $\{ J_i \}_{i \in I}$ is transitive. In light of (\ref{ambiguity}),
this entails that the adjoint action of $\Js$ upon $\irnet$ is also
transitive. Hence, for every $i \in I$ there exists a $g_i \in \Js$ such that 
\begin{equation}
\Rs_i = g_i \Rs_{i_0} g_i{}^{-1} = \Rs_{\tau_{g_i}(i_0)} \quad . \label{free}
\end{equation}
By the Standing Assumptions, this implies $i = \tau_{g_i}(i_0)$, for
every $i \in I$. Of course, for fixed $i \in I$ the group element 
$g_i$ is not unique --- it is determined only up to an element 
of the subgroup of $\Js$ which leaves the algebra $\Rs_{i_0}$ fixed, 
{\it i.e.} the commutator subgroup of $J_{i_0}$, which in our construction
is also the subgroup of $\Ps_+$ leaving the line $l_{i_0}$ fixed. 
$g_i$ itself can be expressed as a product of a finite number of elements in 
$\{ J_i \}_{i \in I}$, and $g_i$ acts adjointly upon our model of 
Minkowski 3-space as the product of the corresponding reflections.

     Let $g\chi(i_0) \equiv \{ gPg^{-1} : P \in \chi(i_0)\}$ denote the 
image under $g \in \Js$ of the wedge $\chi(i_0)$; $g\chi(i_0)$ is itself 
a wedge. For each $i \in I$ and a particular choice of $g_i$ as above, define 
$\chi(i) = \chi(\tau_{g_i}(i_0))$ to be $g_i \chi(i_0)$. Note that
$g_i \chi(i_0)$ is independent of the choice of $g_i \in \Js$ satisfying
equation (\ref{free}), since any element of $\Ps_+$ leaving $l_{i_0}$
fixed also leaves $\chi(i_0)$ fixed. 
One then has
\begin{equation}
J_i \chi(j) = J_i g_j \chi(i_0) = \chi(\tau_{J_i g_j}(i_0)) =
\chi(\tau_i(\tau_{g_j}(i_0))) = \chi(\tau_i(j)) \quad . \label{cov}
\end{equation}

     Since the results of Section 2 entail that $\Js$ is isomorphic to
the proper Poincar\'e group $\Ps_+$, $\Js$ contains an (anti-)unitary
representation $U(\Ps_+)$ of $\Ps_+$; indeed, from the results of
Section 2, one actually has $\Js = U(\Ps_+)$. We define the algebra
$\Rs(\chi(i))$ corresponding to the wedge $\chi(i)$ to be $\Rs_i$. 
Using (\ref{cov}), one then finds that
$$J_i \Rs(\chi(j)) J_i = J_i \Rs_j J_i = \Rs_{\tau_i(j)} = \Rs(\chi(\tau_i(j)))
= \Rs(J_i \chi(j)) \quad   ,  $$
for every $i,j \in I$. This implies that the net $\{ \Rs(\chi(i)) \}_{i \in I}$
is covariant under the representation $U(\Ps_+)$. In addition,
since by construction $J_i \chi(i) = \chi(i)'$ ($J_i$ is the reflection about
the line $l_i = J_i$ and the edge of the wedge $\chi(i)$ is 
$g_i J_{i_0} g_i{}^{-1} = J_i$), it follows that one has
$$\Rs(\chi(i))' = \Rs_i{}' = J_i \Rs_i J_i = J_i \Rs(\chi(i)) J_i = 
\Rs(J_i \chi(i)) = \Rs(\chi(i)') \quad .         $$
This is the property known as Haag duality in AQFT.

     We have therefore proven the following theorem.

\begin{theorem} \label{mainqft} Let $\omega$ and $\inet$ satisfy the
Standing Assumptions, and let the pair $(\Js,\{ J_i \}_{i \in I})$
fulfill Axioms A.1 -- A.6 in Section 2.1. Then there exists a
bijection $\chi : I \rightarrow \Ws$ such that 
$J_i \chi(j) = \chi(\tau_i(j))$, for all $i,j \in I$. The group $\Js$ 
forms an (anti-)unitary representation of the proper Poincar\'e group $\Ps_+$, 
wherein each $J_i$ represents the reflection about the spacelike line in 
$\RR^3$ which is the edge of the wedge $\chi(i)$.  Moreover, the bijection
$\chi$ can be chosen so that with $\Rs(\chi(i)) \equiv \Rs_{i}$, the
collection $\{ \Rs(\chi(i)) \}_{i \in I}$ forms a collection of von
Neumann algebras satisfying Haag duality which is covariant under this
representation of $\Ps_+$. 
\end{theorem}

     We have at present no proof that the bijection $\chi$ can be 
selected in such a manner that the resulting set
$\{ \Rs(\chi(i)) \}_{i \in I}$ satisfies isotony. This is because
$\{ \Rs(\chi(i)) \}_{i \in I}$ is isotonous if and only if
the map $\chi : I \rightarrow \Ws$ is order-preserving, and we
do not know how to assure this. If the order structure on $I$ expresses
the ordering of the laboratories discussed above, then unless such a 
choice of $\chi$ can be made, the conceptual problem at hand has not yet been
satisfactorily solved. We feel it likely that the 
group $\Js$ can only satisfy all the assumptions A1 -- A6 if the
order structure on $I$ is consistent with the order structure
on $\{ \chi(i) \}_{i \in I}$, but a proof to this effect would
require a development of modular theory in a direction which has
hardly been considered in the literature. In anticipation of such 
a future theory, we can at least prove the following theorem.

\begin{theorem} \label{mainqft2} Let $\omega$ and $\inet$ satisfy the
Standing Assumptions, and let the pair $(\Js,\{ J_i \}_{i \in I})$
fulfill Axioms A.1 -- A.6 in Section 2.1. If the bijection 
$\chi : I \rightarrow \Ws$ from Theorem \ref{mainqft} is order-preserving, 
then $\{ \Rs(\chi(i)) \}_{i \in I}$ is a net of von Neumann algebras 
satisfying locality. Furthermore, the CGMA is satisfied by the pair 
$(\omega,\{ \Rs(\chi(i)) \}_{i \in I})$.
\end{theorem}

\begin{proof} In light of the assumed isotony of 
$\{ \Rs(\chi(i)) \}_{i \in I}$ and the Haag duality from 
Theorem \ref{mainqft}, it follows that if 
$\chi(j) \subset \chi(i)'$, then 
$\Rs(\chi(j)) \subset \Rs(\chi(i)') = \Rs(\chi(i))'$, {\it i.e.} the
net is local. The fact that the CGMA is
satisfied by $(\omega,\{ \Rs(\chi(i)) \}_{i \in I})$ is a trivial
consequence of the construction and isotony.
\end{proof}

     Once one has such a net of wedge algebras, it is then standard
\cite{BW} to generate a maximal local net $\rnet$ which also contains
algebras of observables localized in compact regions and is
Poincar\'e covariant under $U(\Ps_+)$.  For the sake of completeness,
we mention that there exist examples of quantum fields, {\it e.g.} the
free scalar Bose field on three-dimensional Minkowski space in the
vacuum state, which fulfill the hypotheses of Theorems \ref{mainqft}
and \ref{mainqft2} \cite{Wh}.
 
     An immediate consequence of these theorems is that the
modular symmetry group $\Js$ must also contain a strongly continuous
unitary representation of the time translation subgroup of $\Ps_+$,
which is usually interpreted as describing the dynamics of the
covariant quantum system. Hence, also the dynamics of the quantum
system is determined by $\omega$ and $\inet$, under the stated 
conditions.

     In \cite{BDFS} a purely algebraic stability condition called the
\msc \ was identified for reasons we shall not explain here.  The \msc
\ requires that the modular unitaries
$\Delta_j{}^{it}$ associated with $(\Omega,\Rs_j)$ in Tomita--Takesaki
theory are contained in $\Js$, for each $j \in I$ and $t \in \RR$.

\begin{cor} If, in addition to the hypotheses of Theorem 
\ref{mainqft2}, the pair $(\omega,\inet)$ fulfills the \msc,
then the modular unitaries $\Delta_{\chi(j)}{}^{it}$ associated with
$(\Omega,\Rs(\chi(j)))$ are contained in $U(\Pid) \subset \Js$, 
for each $j \in I$ and $t \in \RR$, and represent the boost group
leaving the wedge $\chi(j)$ invariant.\footnote{This property is 
called modular covariance in the AQFT literature.}
\end{cor}

\begin{proof} In Section 5.2 of \cite{BDFS} are given sufficient 
conditions for the adjoint action of the modular unitaries
$\Delta_j^{it}$ upon the net constructed above to be implemented by
the specified Poincar\'e transformations. All of those conditions are
satisfied here with the possible exception that the adjoint action of
all modular unitaries may not act transitively upon the net and that
the conditions (ii) and (iii) of the CGMA in \cite{BDFS} may not
hold. However, in \cite{BFS2,F} it is shown that the former assumption
may be dropped. Furthermore, the role of conditions (ii) and (iii) in
\cite{BDFS} was to assure that the adjoint action of modular objects
was implemented by Poincar\'e transformations --- but this is assured
here by the construction above and the \msc. Hence, here these
assumptions may also be dropped.
\end{proof}

\section{Conclusion and Outlook}

     We have shown that it is possible to derive a space--time from
the operationally motivated quantum data of a state modelling the
preparation of the quantum system and a net of algebras of the
observables of the quantum system which is indexed in some suitable
manner, {\it e.g.} by the laboratories in which the observables were
measured. This has been done for the simplest nontrivial case ---
three-dimensional Minkowski space.  A similar derivation of
four-dimensional Minkowski space has been made in \cite{Wh}, though
surely not in the optimal manner. Although it is not yet clear which
class of space--times could be attainable through this approach,
it is likely that one could at least be able to derive in this manner
all space--times with a sufficiently large isometry group.

     We mentioned above the likelihood that a given net $\inet$ will
not satisfy the Standing Assumptions and that it may be necessary to
pass to a subnet. This is because experience has shown that only the
modular involutions associated with algebras localized in certain
types of regions will have a suitable adjoint action upon the net -
cf. \cite{BDFS} for a discussion of this matter.  However, a subnet of
the original net may well satisfy the Standing Assumptions. From this
point of view, the results of \cite{BW} assert that in any finite
component quantum field theory on Minkowski space satisfying the
Wightman axioms to which can be locally associated a net $\rnet$ of
local von Neumann algebras, there always exists a subnet $\wrnet$
which satisfies our Standing Assumptions. Hence, given a net $\inet$,
one would proceed to a subnet satisfying the Standing Assumptions and
suitable absolute geometric axioms, construct the space--time,
make a suitable identification between the elements of the subnet and
algebras associated with special regions of the space--time, then
attempt to use the inclusion relations in the original net $\inet$ to
identify the remaining algebras in $\inet$ with algebras associated
with suitable regions in the derived space--time.

     In Theorem \ref{mainqft} the crucial hypothesis that Axioms A.1
-- A.6 are satisfied is imposed upon the auxiliary (and
non-operational) object $\Js$. And in order to obtain Theorem
\ref{mainqft2} it was necessary for us to posit that $\chi$ was
order-preserving. It would be desirable to determine conditions upon
the net $\inet$ directly which would, by some suitable extension of
the current state of modular theory, imply that the group $\Js$
fulfills the said axioms and that $\chi$ is order-preserving. This
will involve making progress in a field of mathematics which is yet in
its infancy. The modular theory of Tomita--Takesaki was initially
formulated and developed for a state on a single algebra. Only
relatively recently, particularly motivated by questions in AQFT, have
researchers considered a state on a pair of algebras and studied
relations among the modular objects implied by relations between the
algebras. A notable extension of the theory to more than two algebras
can be found in the recent papers \cite{Wies,KW} (though the essential
nature of the insights there were for pairs of algebras, as well).  To
attain a theorem of the type we would like to see some day, it will be
necessary to develop the modular theory of a state and a net of
algebras. Such a theory would have many applications besides the one
we envision.

\bigskip

\noindent {\bf Acknowledgements}:  The authors profitted from discussions
with Prof. Detlev Buchholz. We also wish to thank one of the anonymous
referees, whose advice led us to a simplification of the axiom system. \\


\begin{thebibliography}{Bor}
\bibitem{Ah}J. Ahrens, Begr\"undung der absoluten Geometrie des Raumes aus dem
Spiegelungsbegriff, {\sl Math. Zeitschr., \bf 71}, 154--185 (1959).
\bibitem{Au}J. Audretsch, Riemannian structure of space--time as a 
consequence of quantum mechanics, {\sl Phys. Rev., \bf D27}, 2872--2884
(1983).
\bibitem{AL}J. Audretsch and C. L\"ammerzahl, The conformal structure
of space--time in a constructive axiomatics based on elements of
quantum mechanics, {\sl Gen. Rel. Grav., \bf 27}, 233--246 (1995).
\bibitem{Ba}F. Bachmann, {\it Aufbau der Geometrie aus dem Spiegelungsbegriff},
second edition, Springer-Verlag, Berlin, New York, 1973.
\bibitem{BBPW}F. Bachmann, A. Baur, W. Pejas and H. Wolff, Absolute
geometry, in: {\it Fundamentals of Mathematics}, Vol. 2, MIT Press, 
Cambridge, Mass., 1986.
\bibitem{Ban}U. Bannier, Intrinsic algebraic characterization of space-time 
structure, {\sl Int. J. Theor. Phys., \bf 33}, 1797--1809 (1994). 
\bibitem{BHF}U. Bannier, R. Haag and K. Fredenhagen, Structural definition of
space-time in quantum field theory, unpublished preprint, 1989.
\bibitem{BL}A. Bauer and R. Lingenberg, Affine and projective planes, in:
{\it Fundamentals of Mathematics}, Vol. 2, MIT Press, Cambridge, Mass., 1986.
\bibitem{BW}J. Bisognano and E.H. Wichmann, On the duality condition for a 
hermitian scalar field, {\sl J. Math. Phys., \bf 16}, 985--1007 (1975).
\bibitem{Bor0}H.-J. Borchers, On modular inclusion and spectrum condition, 
{\sl Lett. Math. Phys., \bf 27}, 311--324 (1993).
\bibitem{Bor}H.-J. Borchers, On revolutionizing quantum field theory with 
Tomita's modular theory, {\sl J. Math. Phys., \bf 41}, 3604--3673 (2000).
\bibitem{BR}O. Bratteli and D.W. Robinson, {\it Operator Algebras and
Quantum Statistical Mechanics I}, Springer-Verlag, Berlin, Heidelberg, 
New York, 1979.
\bibitem{BS}D. Buchholz and S.J. Summers, An algebraic characterization of 
vacuum states in Minkowski space, {\sl Commun. Math. Phys., \bf 155}, 
449--458 (1993).
\bibitem{BFS1}D. Buchholz, M. Florig and S.J. Summers, An algebraic 
characterization of vacuum states in Minkowski space, II, {\sl Lett. Math.
Phys., \bf 49}, 337--350 (1999).
\bibitem{BDFS}D. Buchholz, O. Dreyer, M. Florig and S.J. Summers, Geometric 
modular action and spacetime symmetry groups, {\sl Rev. Math. Phys., \bf 12}, 
475--560 (2000).
\bibitem{BFS3}D. Buchholz, M. Florig and S.J. Summers, The second law of 
thermodynamics, TCP, and Einstein causality in anti-de Sitter space-time, 
{\sl Class. Quant. Grav., \bf 17}, L31--L37 (2000).
\bibitem{BMS}D. Buchholz, J. Mund and S.J. Summers, Transplantation of local 
nets and geometric modular action on Robertson-Walker space-times, {\sl Fields 
Inst. Commun., \bf 30}, 65--81 (2001).
\bibitem{BS2}D. Buchholz and S.J. Summers, An algebraic 
characterization of vacuum states in Minkowski space, III: Poincar\'e
covariance, manuscript in preparation.
\bibitem{BFS2}D. Buchholz and S.J. Summers, Geometric modular action and 
modular covariance, manuscript in preparation.
\bibitem{BFS4}D. Buchholz and S.J. Summers, The second law of 
thermodynamics and vacuum states on anti-de Sitter space--time, manuscript
in preparation.
\bibitem{BK}H. Busemann and P. Kelly, {\it Projective Geometry and 
Projective Metrics}, Academic Press, New York, 1953.
\bibitem{Cox1}H.S.M. Coxeter, A geometrical background for DeSitter's
world, {\sl Amer. Math. Monthly, \bf 50}, 217--228 (1943).
\bibitem{Cox2}H.S.M. Coxeter, {\it Introduction to Geometry}, Wiley and
Sons, Inc., New York, 1961.
\bibitem{Cox3}H.S.M. Coxeter, {\it Non-Euclidean Geometry}, fourth edition, 
University of Toronto Press, Toronto, 1961.
\bibitem{Cox4}H.S.M. Coxeter, {\it Projective Geometry}, second edition,
Springer-Verlag, Berlin, New York, 1987.
\bibitem{EPS}J. Ehlers, F.A.E. Pirani and A. Schild, The geometry of
free fall and light propagation, in: {\it General Relativity}, edited
by L. O'Raifeartaigh, Clarendon Press, Oxford, 1972.
\bibitem{F}M. Florig, {\it Geometric Modular Action}, Ph.D. Dissertation, 
University of Florida, 1999.
\bibitem{Fr}K. Fredenhagen, Global observables in local quantum physics, in:
{\it Quantum and Non-Commutative Analysis}, Kluwer Academic Publishers, 
Amsterdam, 1993. 
\bibitem{Haag}R. Haag, {\it Local Quantum Physics}, Springer-Verlag, Berlin,  
1992. (A second edition was released in 1996.)
\bibitem{KR}R.V. Kadison and J.R. Ringrose, {\it Fundamentals of the 
Theory of Operator Algebras}, Volume II, Academic Press, Orlando, 1986.
\bibitem{KW}R. K\"ahler and H.-W. Wiesbrock, Modular theory and the 
reconstruction of four--dimensional quantum field theories, {\sl J. Math. 
Phys., \bf 42}, 74--86 (2001).
\bibitem{KO}B. Klotzek and R. Ottenberg, Pseudoeuklidische R\"aume im 
Aufbau der Geometrie aus dem Spiegelungsbegriff, {\sl Zeitschr. f. math. 
Logik und Grundlagen d. Math., \bf 26}, 145--164 (1980).
\bibitem{Sum}S.J. Summers, Geometric modular action and transformation groups,
{\sl Ann. Inst. Henri Poincar\'e, \bf 64}, 409--432 (1996).
\bibitem{Tak}M. Takesaki, {\it Tomita's Theory of Modular Hilbert Algebras and 
Its Applications}, Lecture Notes in Mathematics, Vol. 128, Springer-Verlag,
Berlin, Heidelberg and New York, 1970.                       
\bibitem{Wh}R. White, {\it An Algebraic Characterization of Minkowski Space}, 
Ph.D. Dissertation, University of Florida, 2001.
\bibitem{Wies}H.-W. Wiesbrock, Modular intersections of von Neumann algebras in
quantum field theory, {\sl Commun. Math. Phys., \bf 193}, 269-285 (1998).
\bibitem{Wo}H. Wolff, Minkowskische und absolute Geometrie, I, {\sl
Math. Ann., \bf 171}, 144--164 (1967).
\end{thebibliography}
\end{document}